\documentclass[preprint,aps,showpacs,nofootinbib,preprintnumbers,amsmath,amssymb]{revtex4-1}
\usepackage{}
\usepackage{epsfig}
\usepackage{dcolumn}
\usepackage{bm}
\usepackage[usenames ,dvipsnames]{xcolor}
\usepackage{slashed}

\usepackage{graphicx,color}

\begin{document}
\title{A $Z^\prime$ Model for $b\to s \ell\bar \ell$ Flavour Anomalies}

\author{Cheng-Wei Chiang$^{1,2,3}$\footnote{Electronic address: chengwei@ncu.edu.tw}, Xiao-Gang He$^{4,3,5}$\footnote{Electronic address: hexg@phys.ntu.edu.tw}, German Valencia$^{6}$\footnote{Electronic address: German.Valencia@monash.edu.au }}
\affiliation{
$^{1}$Department of Physics and Center for Mathematics and Theoretical Physics,
National Central University, Chungli, Taiwan 32001\\
$^{2}$Institute of Physics, Academia Sinica, Taipei, Taiwan 11529\\
$^{3}$Physics Division, National Center for Theoretical Sciences, Hsinchu, Taiwan 30013\\
$^{4}$INPAC,Department of Physics and Astronomy, Shanghai Jiao Tong University, Shanghai\\
$^{5}$Department of Physics, National Taiwan University, Taipei, \\
$^{6}$School of Physics and Astronomy, Monash University, 3800 Melbourne Australia.\footnote{On leave from Department of Physics, Iowa State University, Ames, IA 50011.}
}

\begin{abstract}
We study the implications of flavour-changing neutral currents (FCNC's) in a model with the $SU(2)_l\times SU(2)_h\times U(1)_Y$ electroweak gauge symmetry for several anomalies appearing in $b\to s \ell\bar \ell$ induced $B$ decays in LHCb data.  In this model, $SU(2)_l$ and $SU(2)_h$ govern the left-handed fermions in the first two generations and the third generation, respectively.  The physical $Z$ and $Z'$ generate the $b\to s$ transition at tree level, leading to additional contributions to the $b \to s$ semileptonic operators ${\cal O}_{9,10}$.  We find that although $B_s$-$\bar B_s$ mixing constrains the parameters severely, the model can produce values of ${\cal C}^{\rm NP}_{9,10}$ in the range determined by Descotes-Genon {\it et. al.} in Ref.~\cite{Descotes-Genon:2015uva} for this scenario to improve the global fit of observables in decays induced by the $b\to s \mu \bar \mu$ transition.  The $Z'$ boson in this model also generates tree-level FCNC's for the leptonic interactions that can accommodate the experimental central value of $R_K = {\cal B}(B\to K \mu \bar \mu)/{\cal B}(B\to K e\bar e)=0.75$. In this case, the model predicts sizeable branching ratios for $B\to K e \bar \tau$, $B\to K \tau \bar e$, and an enhancement of $B\to K \tau \bar \tau$ with respect to its SM value.

\end{abstract}
\maketitle


\section{Introduction}

Experimental data have hinted at several anomalies in $B$ decays induced by the flavour-changing neutral current (FCNC) process $b\to s \ell \bar \ell$. In 2013, LHCb measured four observables related to the angular distribution of $B\to K^*\mu^+ \mu^-$ in six bins of dimuon invariant mass squared, $q^2$, and found a deviation at the $3.7\sigma$ level from the standard model (SM) in one of them~\cite{Aaij:2013qta}. LHCb also measured the rates for the $B\to K^{(*)}\mu^+ \mu^-$ decay~\cite{Aaij:2014pli}, finding values slightly below the SM expectations.  Recently, with finer binning, LHCb confirmed their earlier anomaly in the angular distribution of $B\to K^*\mu^+ \mu^-$ decay~\cite{Aaij:2015oid}.  In addition, LHCb has studied other modes induced by the $b\to s \mu^+  \mu^-$ transition as well, namely, the $B_s\to \phi\mu^+\mu^-$ decay~\cite{Aaij:2013aln} and also $b\to s e^+ e^-$ in  the $B\to K^* e^+e^-$ decay~\cite{Aaij:2015dea}  with results consistent with the SM. 

A particularly interesting discrepancy between experiment and the SM is in the ratio $R_K$ of the branching fraction of $B^+\to K^+ \mu^+ \mu^-$ to that of $B^+\to K^+ e^+ e^-$. Lepton-universality in the SM predicts $R_K$ to be very close to 1. Yet  LHCb found $R_K \equiv {\cal B}(B\to K \mu \bar \mu)/{\cal B}(B\to K e\bar e) = 0.745^{+0.090}_{-0.074}\pm 0.036$~\cite{Aaij:2014ora} for the dilepton invariant mass squared range of $1 - 6$ GeV$^2$.  This disagreement occurs only at the $2.6\sigma$ level, but would be extremely interesting if confirmed.  

As expected, the anomalies in the $b\to s \ell \bar \ell$ measurements have received considerable attention in the literature \cite{Matias:2012xw} and several models have been put forward as possible new physics explanations~\cite{models}. It has also been argued that more careful treatment of long distance physics would eliminate most of these anomalies, as done most recently in Ref.~\cite{Ciuchini:2015qxb}. Models have also been put forth attempting to explain the apparent lepton non-universality observed in $R_K$ \cite{modelsrk}. A recent analysis of these experimental results is that of Ref.~\cite{Descotes-Genon:2015uva}, where global fits of the observables in terms of new physics parametrised by deviations from the SM values of certain Wilson coefficients are presented. This model-independent analysis and its results are the starting point of our discussions.

In this paper we will focus our discussion around the scenario in which new physics affects primarily the ${\cal C}_9$ and ${\cal C}_{10}$ Wilson coefficients, which has been found in Ref.~\cite{Descotes-Genon:2015uva} to significantly improve the agreement between the measurements and the theoretical predictions. We recall that these coefficients appear in the low-energy effective Hamiltonian responsible for $b\to s \ell \bar \ell$ transitions as follows:
\begin{eqnarray}
&&
{\cal H}_{\rm eff}
= -{4G_F\over \sqrt{2}}V_{tb}V^*_{ts}
\left( {\cal C}^{\ell\ell}_9 {\cal O}_9 + {\cal C}^{\ell\ell}_{10}{\cal O}_{10} \right)\;,\nonumber\\
&&
{\cal O}_9 = {e^2\over 16\pi^2} \left( \bar s \gamma_\mu P_Lb \right)
\left( \bar \ell \gamma^\mu \ell \right) \;,\;\;
{\cal O}_{10} = {e^2\over 16\pi^2} \left( \bar s \gamma_\mu P_Lb \right)
\left( \bar \ell \gamma^\mu\gamma_5 \ell \right) \;,
\end{eqnarray}
where $P_L = (1-\gamma_5)/2$ and, in the absence of flavour universality, ${\cal C}^{\ell\ell}_{9,10}$ can have different values for different lepton flavours.

Within the SM, ${\cal C}^{\ell\ell}_{9,10}$ are approximately the same for all leptons with
${\cal C}_9^{\rm SM} \approx 4.1$, and ${\cal C}_{10}^{\rm SM} \approx -4.1$. To reduce the tension in the global fit associated with the $b\to s \mu \bar \mu$ anomalies, 
the new physics contribution ${\cal C}^{{\rm NP},\mu\mu}_{9}$ is required to be of order $-1.0$ and for scenarios where ${\cal C}^{{\rm NP},\mu\mu}_{10}$ is also not zero, the best fit occurs for ${\cal C}^{{\rm NP},\mu\mu}_{10}\sim 0.3$~\cite{Descotes-Genon:2015uva}. 
To address the anomaly in the value of $R_K$, the absolute value of ${\cal C}^{\rm SM}_{9,10}+{\cal C}^{{\rm NP},ee}_{9,10}$ is required to be larger than that of ${\cal C}^{\rm SM}_{9,10}+{\cal C}^{{\rm NP},\mu\mu}_{9,10}$. 

When going beyond the SM, additional operators with different chiral structures that contribute to $b\to s \ell\bar \ell$ can also be generated, such as
${\cal O}'_9 = (e^2/ 16\pi^2) \left( \bar s \gamma_\mu P_Rb \right) \left( \bar \ell \gamma^\mu \ell \right)$ and ${\cal O}'_{10} = (e^2/16\pi^2) \left( \bar s \gamma_\mu P_Rb \right) \left( \bar \ell \gamma^\mu\gamma_5 \ell \right)$, where $P_R = (1 + \gamma_5)/2$. For the remaining of this paper, we will neglect this possibility and concentrate on a scenario with modified ${\cal C}_{9,10}$ only, corresponding to a particular $Z'$ interpretation of the anomalies. This particular interpretation is motivated by the possibility of lepton non-universality hinted at by $R_K$, and its occurrence in non-universal $Z'$ models that single out the third generation.

A common extension of the SM that produces tree-level FCNC's is a $Z'$ boson, particularly when it is non-universal in generations. This new interaction can have different types of chiral structures in both quark and lepton sectors. A model that singles out the third generation with an additional right-handed interaction~\cite{us} leads to tree-level FCNC's for ${\cal O}'_{9,10}$ which, according to the global fits of Ref.~\cite{Descotes-Genon:2015uva}, do not help much in addressing the observed anomalies. At one-loop level, it is possible to produce the pattern ${\cal C}_9^{\rm NP}={\cal C}_{10}^{\rm NP}$ which is disfavoured by the data on $B_s\to \mu\mu$. In this context, a model that more naturally fits the ${\cal C}_{9,10}$ scenario is one where the $SU(2)_L$ gauge group in the SM is extended to be generation-dependent~\cite{Ma:1988dn}, an example of which has been dubbed `top-flavour' before~\cite{Muller:1996dj}. 

The model has the $SU(2)_l\times SU(2)_h\times U(1)_Y$ gauge symmetry, where $SU(2)_l$ governs the  left-handed fermions in the first two light generations and $SU(2)_h$ governs those in the third heavy generation. This model has been studied before by two of us in Ref.~\cite{Chiang:2009kb}. It affects the $b\to s \ell\bar \ell$ process at tree level mostly through modifications to ${\cal C}_{9,10}$. The relevant parameters are severely constrained by $B_s$-$\bar B_s$ mixing.  Nevertheless, the model can still produce values of ${\cal C}_{9,10}^{\rm NP}$ in the right ranges to improve the global fits as described in Ref.~\cite{Descotes-Genon:2015uva}. In addition, the model can break lepton universality and lepton number, accommodating $R_K$ and predicting sizeable branching ratios for $B \to K e \bar \tau$, $B\to K \tau \bar e$ and an enhancement of $B\to K \tau \bar \tau$ with respect to its SM value.

This paper is organized as follows.  In Section~\ref{sec:FCNC}, we review the tree-level FCNC's induced by the $Z$ and $Z'$ exchanges in the model, deriving the basis for the latter analyses.  In Section~\ref{sec:bsll}, we compute the corrections to the Wilson coefficients ${\cal C}_{9,10}^{NP}$ occurring in this model.  In Section~\ref{sec:numbers}, we update the global fit to the electroweak precision data and the $B_s$-$\bar B_s$ mixing constraint, thereby obtaining preferred ranges of the theory parameters.  The results are then used to evaluate ${\cal C}_{9,10}^{\rm NP}$ numerically and to check against the preferred values presented in Ref.~\cite{Descotes-Genon:2015uva}.  Taking a step further, we make predictions for $R_K$ and the decay branching ratios of $B\to K e \bar \tau$, $B\to K \tau \bar e$, and $B\to K \tau \bar \tau$.  Section~\ref{sec:summary} summarizes our findings.

\section{Tree-level FCNC's due to $Z$ and $Z'$ in the Model \label{sec:FCNC}}

With the gauge group extended from $SU(3)_C\times SU(2)_L \times U(1)_Y$ to $SU(3)_C\times SU(2)_l\times SU(2)_h\times U(1)_Y$, there are additional gauge bosons: a pair of $W'^{\pm}_\mu$ bosons and a $Z'$ boson. With an appropriate Higgs sector, the $SU(2)_l\times SU(2)_h$ symmetry is broken down to $SU(2)_L$ at the TeV scale, leaving the SM gauge group followed by the standard electroweak symmetry breakdown~\cite{Chiang:2009kb} .  The $Z$ and $Z'$ FCNC's relevant to the $b\to s\ell \bar \ell$ transitions are caused by the neutral gauge boson interactions with fermions. 

The left-handed quark doublets $Q_L$, the right-handed quark singlets $U_R$ and $D_R$, the left-handed lepton doublets $L_L$, and the right-handed charged leptons $E_R$ transform under the original gauge group as
\begin{eqnarray}
&&Q^{1,2}_L: (3,2,1,1/3)\;,\;\;Q_L^3: (3,1,2,1/3)\;,\;\;U^{1,2,3}_R: (3,1,1,4/3)\;,\;\;D_R^{1,2,3}: (3,1,1,-2/3)\;,\nonumber\\
&&L^{1,2}_L: (1,2,1,-1)\;,\;\;L_L^3: (1,1,2,-1)\;,\;\;E^{1,2,3}_R: (1,1,1,-2)\;,
\end{eqnarray}
where the numbers in each bracket are the quantum numbers of the corresponding field under $SU(3)_C$, $SU(2)_l$, $SU(2)_h$ and $U(1)_Y$, respectively. The superscript on each field labels the generation of the fermion.

The neutral gauge boson interactions with fermions are given by
\begin{eqnarray}
{\cal L} = \bar\psi \gamma_\mu 
\left[ e A^\mu Q 
+ {g\over c_W} Z_L^\mu \left( T^l_3 + T^h_3 - Qs^2_W \right) 
+ g Z^\mu_H \left( {s_E\over c_E}T^l_3 - {c_E\over s_E}T^h_3 \right) \right ]
\psi\;,
\end{eqnarray}
where $\psi$ represents a quark or lepton field, $T^{l,h}_3$ are the third components of the $SU(2)_{l,h}$ generators, the electric charge $Q$ is given by $Q=T_3+Y/2$ with $T_3 = T^l_3 +T^h_3$, and $s_E$ and $c_E$ respectively are defined in terms of the gauge couplings $g_{1,2}$ of $SU(2)_{l,h}$ by
\begin{align}
s^2_E \equiv \sin^2\theta_E = {g^2_1\over g^2_1+g^2_2}\;\;\;
c^2_E \equiv \cos^2\theta_E = {g^2_2\over g^2_1+g^2_2}\;.
\end{align}
The SM couplings $g$ and $e$ are then given in terms of $g_{1,2}$ and $U(1)_Y$ coupling $g'$  by
\begin{eqnarray}
g^2 = {g^2_1g^2_2\over g^2_1 + g^2_2} \;,\;\;e^2 = {g^2g'^2\over g^2 +g'^2}\;.
\end{eqnarray}

The fields $A,\;Z_{L},\;Z_H$ are defined in terms of the third components $W^3_{l,h}$ of the $SU(2)_{l,h}$ gauge fields and the $U(1)_Y$ gauge field $B$ through the  following transformation:
\begin{eqnarray}
&&\left ( \begin{array}{l}
W^l_3\\W^h_3\\B 
\end{array}
\right )
=
\left (
\begin{array}{rrr}
s_E&c_Ec_W&c_E s_W\\
-c_E&s_Ec_W&s_Es_W\\
0&-s_W&c_W
\end{array}
\right )
\left ( \begin{array}{l}
Z_H\\Z_L\\A 
\end{array}
\right )\;,
\end{eqnarray}
where 
\begin{eqnarray}
&&\;\;s^2_W = {g'^2\over g^2+g'^2}\;\;\;c^2_W = {g^2\over g^2+g'^2}\;.
\end{eqnarray}
In general $Z_{L,H}$ are not mass eigenstates. Writing them in terms of light and heavy mass eigenstates $Z_l$ and $Z_h$, we have
\begin{eqnarray}
Z_L = -\sin\xi Z_h + \cos\xi Z_l\;,\;\;Z_H = \cos\xi Z_h + \sin\xi Z_l\;,
\end{eqnarray}
where a rotation angle $\xi$ is introduced.

Assume that the breaking of  $SU(2)_l\times SU(2)_h$ to $SU(2)_L$ is achieved by a bi-doublet
$\eta: (1,2,2,0)$ with a non-zero vacuum expectation value (VEV), $u \sim {\cal O}({\rm TeV})$, and the subsequent symmetry breaking is achieved by two doublets $\Phi_1: (1,2,1,1)$ and $\Phi_2: (1,1,2,1)$ with respective VEV's $v_1$ and $v_2$ with $v_1^2 + v_2^2 = (174~{\rm GeV})^2$.  We then have to the leading order in $\epsilon \equiv v/u$
\begin{eqnarray}
&&\xi \approx {s_Ec_E\over c_W}(s^2_\beta -s^2_E) \epsilon^2\;,\;\; {m^2_{Z_l}\over m^2_{Z'_h} } \approx \epsilon^2{s^2_Ec^2_E\over c^2_W}\;, 
\end{eqnarray}
where $s^2_\beta \equiv v^2_1/(v^2_1+v^2_2)$.  Because of the mass hierarchy between fermions belonging to the third generation and the first two generations, $s^2_\beta$ is expected to be small.

Now we can express the neutral gauge boson interactions with fermions in the small $\epsilon$ limit as
\begin{eqnarray}
{\cal L} &=& \bar f \gamma_\mu 
\left\{ eQ A_\mu + {g\over c_W} Z^\mu_l \left[ 
T_3 - Qs^2_W - \epsilon^2c^2_E (s^2_E T_3 - T^h_3)
\right] \right. \nonumber\\
&& \left. \qquad
+ g{Z^\mu_h\over s_Ec_E} \left[
s^2_E T_3 -T^h_3 +\epsilon^2 {s^2_E c^4_E\over c^2_W} (T_3 - Qs^2_W)
\right]
\right\} f\;.
\end{eqnarray}
$T^h_3$ acts only on the third generation and the terms proportional to it will induce FCNC's in the fermion mass eigenstate basis.

\section{$b\to s \ell \bar \ell$ transitions \label{sec:bsll}}

Through the exchanges of $Z$ and $Z'$ at tree level, the following effective four-fermion interactions can be induced:
\begin{eqnarray}
{\cal H}_{\rm eff} &=&- {g_Z^2\over 8 m^2_{Z_l}} \epsilon^2 c^2_E 
\left( \bar q \tilde \Delta^{q} \gamma_\mu P_L q \right) 
\left(\bar \ell \gamma^\mu (4 s^2_W - 1 +\gamma_5)\ell \right)
\nonumber\\
&&- {g^2\over 8 s^2_Ec^2_E m^2_{Z_h}}
\left( \bar q \tilde \Delta^q \gamma_\mu P_L q \right) 
\left( \bar \ell \gamma^\mu(s^2_E I - \tilde \Delta^l)(1-\gamma_5)\ell \right)\;,
\end{eqnarray}
where $\tilde \Delta^f= T^\dagger_f {\rm diag}(0,0,1) T_f $ with $\bar f_R M_f f_L = \bar f_R S_f \hat M_f T^\dagger_f f_L$, and $S_f$ and $T_f = (T^f_{ij})$ are unitary matrices for a bi-unitary transformation to obtain the diagonal eigenmass matrix $\hat M_f$. Here we have used the fact that the eigenvalue of $T_3$ for down quarks and charged leptons is $-1/2$. One can further re-write the above expression as
\begin{eqnarray}
{\cal H}_{\rm eff} &=& 
-{4G_F\over \sqrt{2}}V_{tb}V^*_{ts}{\pi\over \alpha}\epsilon^2c^2_E 
{\tilde \Delta^q_{sb}\over V_{tb}V^*_{ts}}\delta_{ij}
\left[ (4s^2_W -1){\cal O}^{ij}_9 +{\cal O}^{ij}_{10} \right]
\nonumber\\
&&-{4G_F\over \sqrt{2}}V_{tb}V^*_{ts}{\pi\over \alpha}\epsilon^2  {\tilde \Delta^q_{sb}\over V_{tb}V^*_{ts}}(s^2_E\delta_{ij} - \tilde \Delta^\ell_{ij})
\left({\cal O}^{ij}_9 - {\cal O}^{ij}_{10} \right)\;,
\label{eqnfcnc}
\end{eqnarray}
where $\tilde \Delta^q_{sb} = T^{q*}_{bs}T^q_{bb}$ and $\tilde \Delta^\ell_{ij} = T^{\ell*}_{3i}T^\ell_{3j}$, and 
\begin{eqnarray}
{\cal O}^{ij}_9 = {e^2\over 16\pi^2}
\left( \bar s \gamma_\mu P_Lb \right) \left( \bar \ell_i \gamma^\mu \ell_j \right) \;,\;\;
{\cal O}^{ij}_{10} = {e^2\over 16\pi^2}
\left( \bar s \gamma_\mu P_Lb \right) \left( \bar \ell_i \gamma^\mu\gamma_5 \ell_j \right)\;.
\end{eqnarray}

From Eq.~(\ref{eqnfcnc}), we can read off the expressions for ${\cal C}^{ij}_{9,10}$ as
\begin{eqnarray}
&&Z\;\;\mbox{contribution}:\;\;{\cal C}^{Z, ij}_9 = {\pi\over \alpha}\epsilon^2c^2_E {\tilde \Delta^q_{sb}\over V_{tb}V^*_{ts}} \left( 4s^2_W - 1 \right)\delta_{ij}\;,\;\;
{\cal C}^{Z, ij}_{10} = {\pi\over \alpha}\epsilon^2c^2_E {\tilde \Delta^q_{sb}\over V_{tb}V^*_{ts}} \delta_{ij}\;,\nonumber\\
&&Z'\;\;\mbox{contribution}:\;\;{\cal C}^{Z', ij}_9 = -{\cal C}^{Z',ij}_{10} = {\pi\over \alpha}\epsilon^2 {\tilde \Delta^q_{sb}\over V_{tb}V^*_{ts}} 
\left( s^2_E\delta_{ij} -\tilde \Delta^\ell_{ij} \right) \;.
\label{coeffs}
\end{eqnarray}
The total new physics contributions to the Wilson coefficients are ${\cal C}^{{\rm NP},ij}_{9,10} = {\cal C}^{Z,ij}_{9,10} + {\cal C}^{Z',ij}_{9,10}$. This implies that within this model and $\tilde\Delta^\ell_{ij}=0$ for $i \not= j$, we have the relation
\begin{eqnarray}
{\cal C}_{10}^{\rm NP}=\frac{{\cal C}_{9}^{\rm NP}}{2s_W^2(\sec2\theta_E+1)-1}.
\label{relation}
\end{eqnarray}

\section{Numerical analysis \label{sec:numbers}}

We now explore the numerical ranges that can be obtained for ${\cal C}^{NP}_{9,10}$ and compare them with those of Ref.~\cite{Descotes-Genon:2015uva} that can reduce the tension between the predictions and measurements for the observables in $b\to s \ell\bar \ell$ induced $B$ decays.  For this purpose, we need to know the constraints for the new model parameters, $\epsilon$, $c_E$, $\tilde \Delta^q_{sb}$ and $\tilde \Delta^\ell_{ij}$.

\begin{figure}[!h]
\centerline{ 
\includegraphics[width=0.6\textwidth]{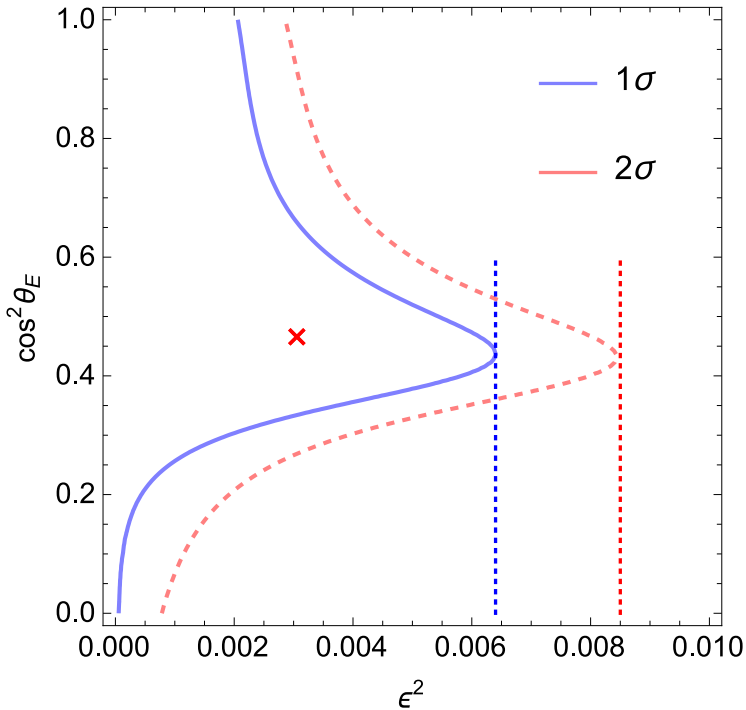} 
}
\caption{$\chi^2$ contours of fit to   electroweak precision data.  The best-fit point, $1\sigma$ contour and $2\sigma$ contour are marked by a red cross, a blue solid curve, and a red dashed curve, respectively.  The vertical dotted lines mark the $1\sigma$ and $2\sigma$ upper bounds on $\epsilon^2$.}
\label{modelp}
\end{figure}

The model parameters $\epsilon$ and $c_E$ were constrained by the electroweak precision data in Ref.~\cite{Chiang:2009kb}.  We update this fit here using the latest data~\cite{Agashe:2014kda}.  The $\chi^2$ contours on the $\epsilon^2$-$c_E^2$ plane are shown in Fig.~\ref{modelp}.  It is seen that the best fit values for $\epsilon^2$ and $c_E^2$ are $0.0031$ and $0.4629$, respectively.  The former indicates that both the VEV of $\eta$ and the $Z'$ mass are about $3$~TeV.  The $1\sigma$ and $2\sigma$ upper bounds on $\epsilon^2$ are $0.0064$ and $0.0085$ as marked by the vertical dashed and dotted lines, respectively.  $c_E^2$ can range from 0 to 1 at both $1\sigma$ and $2\sigma$ levels. In particular, Eq.~(\ref{relation}) allows ${\cal C}_{10}^{NP}$ to be vanishing when $\theta_E=\pm\pi/4$.

The parameters $\tilde \Delta_{ij}^\ell$ involve only leptons and are not well constrained yet.  On the other hand, $\tilde \Delta^q_{sb}$ is severely constrained by the $B_s$-$\bar B_s$ mixing.  The contribution of $Z'$ exchange to $\Delta M_{B_s}$ of the $B_s$ mixing system is given by
\begin{align}
\begin{split}
\Delta M_{B_s}^{Z'}
&= \frac{G_F}{\sqrt{2} m_{B_s}} \epsilon^2 (\tilde \Delta^q_{sb})^2  \langle \bar B_s | 
\left( \bar s \gamma^\mu P_L b \right) \left( \bar s \gamma_\mu P_L b \right) | B_s \rangle \ \hat\eta_{B}
\\
&= \frac{\sqrt{2}G_F}{3} \left(\epsilon \tilde \Delta^q_{sb} \right)^2 
m_{B_s} f^2_{B_s} B_{B_s}\hat\eta_{B}
\\
&= \sqrt{2}G_F\frac{\alpha}{6\pi s_W^2}V_{tb}V^*_{ts} m_{B_s} f^2_{B_s} B_{B_s}\hat\eta_{B} \tilde\Delta^q_{sb}
\left[ {\cal C}_9^{\rm NP} + {\cal C}_{10}^{\rm NP}(1-2s_W^2) \right] ~,
\end{split}
\label{dmbs}
\end{align}
where the last expression has been written in terms of the Wilson coefficients ${\cal C}_{9,10}^{\rm NP}$ given in Eq.~(\ref{coeffs}) to emphasize the correlation.  Note that ${\cal C}_{9,10}^{\rm NP}$ are also linear in the flavour-changing coupling $\tilde\Delta^q_{sb}$.

Numerically,  $f_{B_s}\sqrt{B_{B_s}} = (216\pm 15)$~ MeV~\cite{Artuso:2015swg}.  We have also included the QCD correction factor $\hat\eta_{B}\approx 0.84$~\cite{Artuso:2015swg} to account for the renormalization group running of the operator from the electroweak scale to the $B_s$ scale and neglected a small correction from additional running of the operator between the electroweak scale and the $Z'$ scale. 
For our numerical estimates, it is convenient to rewrite the non-perturbative factors in terms of the SM contribution:
\begin{align}
\begin{split}
\frac{\Delta M_{B_s}^{Z'} }{\Delta M_{B_s}^{\rm SM}}
&=
\frac{2\sqrt{2}\pi^2}{G_FM_W^2S_0[x_t]}
\left(\frac{\epsilon \tilde \Delta^q_{sb}}{|V_{tb}V^*_{ts}|}\right)^2 
\\
&\approx 
161.8 \left(\frac{\epsilon \tilde \Delta^q_{sb}}{|V_{tb}V^*_{ts}|}\right)^2 \left(\frac{2.29}{S_0[x_t]} \right) .
\end{split}
\label{dmbssm}
\end{align}
We thus remove the main uncertainties from non-perturbative QCD factors, and ignore all but parametric uncertainties in the short-distance part due to the $Z^\prime$ exchange.

Experiments have determined $\Delta M_{B_s}$ to high precision.  The latest HFAG average~\cite{Amhis:2014hma} of the CDF~\cite{Abulencia:2006ze} and LHCb~\cite{Aaij:2013mpa} results is $\Delta M_{B_s}^{\rm exp} =(17.757\pm 0.021){\rm ~ps}^{-1} $. This value is consistent with the latest SM prediction, $\Delta M_{B_s}^{\rm SM} = (18.3 \pm 2.7) {\rm~ ps}^{-1}$~\cite{Artuso:2015swg}, leaving little room for new physics, particularly if it interferes constructively with the SM as the term in Eq.~(\ref{dmbs}) does.  Combining these errors in quadrature, we restrict the new physics contribution to be $0\leq \Delta M_{B_s}^{Z'} \leq 2.7 ~(5.4)~{\rm ps}^{-1}$ at $1\sigma$ ($2\sigma$).

\begin{figure}[!h]
\centerline{
\includegraphics[width=0.6\textwidth]{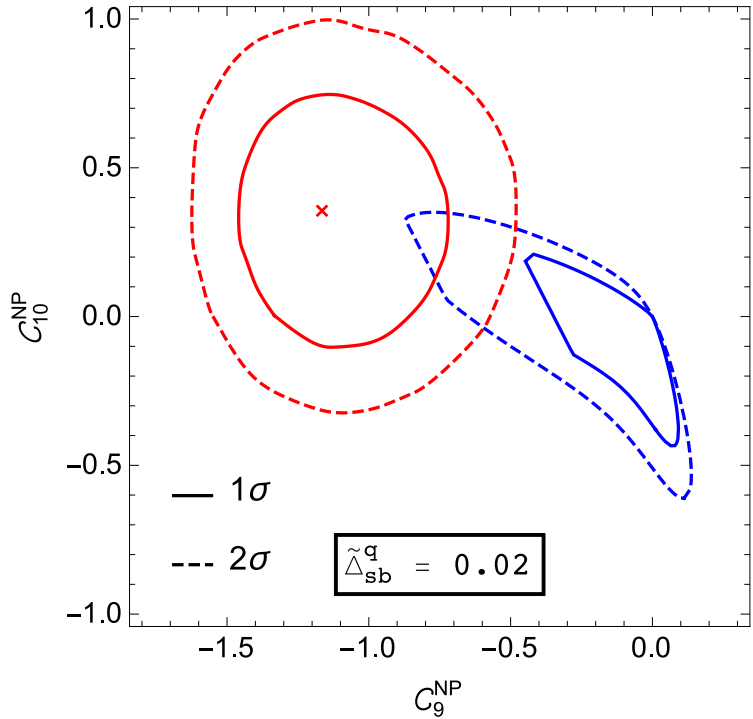}
}
\caption{The region allowed by the electroweak precision data fit and $\Delta M_{B_s}^{\rm exp}$ is shown in blue (the curves on the right).  The region allowed by a global fit to $b\to s\ell\bar\ell$ observables in Ref.~\cite{Descotes-Genon:2015uva} is shown in red (the curves on the left) for comparison.}
\label{c9-10}
\end{figure}

In Fig.~\ref{c9-10}, we show the 1$\sigma$ (solid blue) and 2$\sigma$ (dashed blue) contours in the ${\cal C}^{\rm NP}_{9,10}$ parameter space, as determined by the  electroweak precision data in Fig.~\ref{modelp} and by $\Delta M_{B_s}^{\rm exp}$, for the particular value $\tilde\Delta^q_{sb}=0.02$. This value is chosen so that it allows the $2\sigma$ contour to be in the vicinity of the best fit for the $b\to s\ell\bar\ell$ anomalies in the ${\cal C}^{\rm NP}_{9,10}$ scenario of Ref.~\cite{Descotes-Genon:2015uva}, shown by the red $\times$.  The $1\sigma$ (solid red) and $2\sigma$ (dashed red) contours from that global fit are also shown in the figure.  Our results show that although it is not possible to reach the best-fit point within our model, there is a substantial overlap at the $2\sigma$ level between the values of ${\cal C}^{\rm NP}_{9,10}$ that can be obtained in this model and those that improve the 
$b\to s\ell\bar\ell$ global fit.

So far we have assumed $\tilde \Delta^\ell_{ij}=0$ for $i \not= j$ in Eq.~(\ref{coeffs}).  Nevertheless, they can be non-vanishing and lead to the possibilities of lepton non-universality and of lepton-flavour violation within this model.  To limit the parameter space, we start with a point within the $2\sigma$ contour of Fig.~\ref{c9-10} that is closest to the best-fit point; namely,
\begin{align}
\begin{split}
&
\tilde\Delta^q_{sb}=0.02 ~,\ \epsilon=0.088 ~,\ \cos\theta_E=0.63 ~,
\\
&
\Delta M_{B_s}^{Z'}=5.4 {\rm~ps}^{-1} ~,\  {\cal C}^{\rm NP}_{9}=-0.87 ~,\ {\cal C}^{\rm NP}_{10}=0.32 ~.
\end{split}
\label{bestn}
\end{align}
Since $\tilde \Delta^\ell_{ij} = T^{\ell*}_{3i}T^\ell_{3j}$, setting $\tilde \Delta^\ell_{22} =0$ will maximize ${\cal C}^{{\rm NP},\mu\mu}_9$. This 
implies that for real $T^\ell$, it has the following form
\begin{eqnarray}
T^\ell = \left ( \begin{array}{ccc}
\cos\theta&0&-\sin\theta\\0&1&0\\ \sin\theta&0&\cos\theta
\end{array}
\right )\;.
\end{eqnarray}

The non-zero entries for $\tilde \Delta^\ell_{ij}$ are then: $\tilde \Delta^\ell_{11} = \sin^2\theta$, $\tilde \Delta^\ell_{13} = \tilde \Delta^\ell_{31}= \sin\theta\cos\theta$ and $\tilde \Delta^\ell_{33} = \cos^2\theta$.  Varying the value for $\sin\theta$ will change our predictions for $B\to K (e \bar e, \tau\bar \tau, e\bar \tau, \tau \bar e)$, breaking both lepton universality and lepton flavour conservation.

In particular, the model can accommodate the value  $R_K = 0.745^{+0.090}_{-0.074}\pm 0.036$. 
Neglecting the lepton masses, we have
\begin{align}
\begin{split}
&
{\cal B}(B\to K \ell _i \bar \ell_j) \propto 
\left ({\cal C}^{\rm SM}_9 + {\cal C}^{{\rm NP},ij}_9\right )^2 
+ \left ({\cal C}^{\rm SM}_{10} + {\cal C}^{{\rm NP},ij}_{10}\right )^2 ~,
\\
&
{\cal B}(B\to K e\bar e) \propto 
\left ({\cal C}^{\rm SM}_9 + {\cal C}^{{\rm NP}, \mu\mu}_9-\tilde \Delta^\ell_{11}\frac{{\cal C}^{{\rm NP},\mu\mu}_{10}}{ c^2_E}\right )^2 
+ \left[ {\cal C}^{\rm SM}_{10} + {\cal C}^{{\rm NP}, \mu\mu}_{10}
\left( 1+\frac{\tilde \Delta^\ell_{11}}{c^2_E} \right)
\right]^2 ~,
\\
&
{\cal B}(B\to K \tau \bar \tau) \propto
\left ({\cal C}^{\rm SM}_9 + {\cal C}^{{\rm NP}, \mu\mu}_9-\tilde \Delta^\ell_{33}\frac{{\cal C}^{{\rm NP},\mu\mu}_{10}}{ c^2_E}\right )^2 
+ \left[ {\cal C}^{\rm SM}_{10} + {\cal C}^{{\rm NP}, \mu\mu}_{10}
\left( 1+\frac{\tilde \Delta^\ell_{33}}{c^2_E} \right)
\right]^2 ~,
\\
&
{\cal B}(B\to K e\bar \tau, \tau \bar e) \propto 
2 \left({\cal C}^{{\rm NP}, \mu\mu}_{10}\right)^2
\left(\frac{\tilde \Delta^\ell_{13}}{1-2c_E^2}\right )^2  .
\end{split}
\end{align}
With the numbers given in Eq.~(\ref{bestn}), we then obtain
\begin{align}
&R_K=0.745 
\Rightarrow \sin^2\theta = 0.37 ~,
\\
&\frac{{\cal B}(B\to K \tau \bar \tau)}{{\cal B}(B\to K \mu\bar \mu)} 
= 1.36 ~,
\\
&\frac{{\cal B}(B\to K (e \bar \tau, \tau \bar e))}{{\cal B}(B\to K \mu\bar \mu)} 
= 0.037 ~.
\end{align}

\section{Summary and Conclusions \label{sec:summary}}

As one intriguing feature, the model with the $SU(2)_l\times SU(2)_h\times U(1)_Y$ electroweak gauge symmetry proposed earlier~\cite{Chiang:2009kb} has flavour-changing neutral currents (FCNC's) at tree level, mediated by both $Z$ and $Z'$ bosons.  In this model, fermions of the first two generations and those of the third generations are charged respectively under the $SU(2)_l$ and $SU(2)_h$ groups.  A scalar $\eta$ in the bi-fundamental representation of $SU(2)_l\times SU(2)_h$ is introduced to break the symmetry to $SU(2)_L$ in the standard model (SM) with a vacuum expectation value (VEV) of $u$.  The $SU(2)_L \times U(1)_Y$ is then broken by two Higgs doublets, $\Phi_1$ and $\Phi_2$, with respective VEV's $v_1$ and $v_2$ and $v^2 = v_1^2 + v_2^2 = (174~{\rm GeV})^2$.

In this work, we first extracted two important parameters $\epsilon$ and $\cos\theta_E$ of the model using the latest electroweak precision data, where $\epsilon^2$ denotes the ratio $(v/u)^2$ and $\cos^2\theta_E$ denotes the ratio of the two $SU(2)$ gauge couplings, $g_2^2 / (g_1^2 + g_2^2)$.  Their best-fit values were found to be $0.0031$ and $0.4629$, respectively.  The former indicates that the breaking scale of the $SU(2)_l\times SU(2)_h$ symmetry as well as the $Z'$ mass are both around 3 TeV.

Based on the results of a global fit~\cite{Descotes-Genon:2015uva} to the $b\to s\ell\ell$ anomalies recently reported by LHCb, we discussed how the FCNC interactions in our model would affect the Wilson coefficients ${\cal C}_{9,10}^{\rm NP}$ associated with the $b \to s \ell \bar \ell$ operators ${\cal O}_{9,10}$ to get close to the values found by the global fit to address the anomalies.    We noticed that a stringent constraint on ${\cal C}_{9,10}^{\rm NP}$ came from the $B_s$-$\bar B_s$ mixing data, and showed the correlation within the model.  We found that at the $2\sigma$ level, the model could accommodate the best-fit values for ${\cal C}_{9,10}^{\rm NP}$ while satisfying the  $\Delta M_{B_s}$ measurement.

Moreover, the $Z'$ boson could have non-universal or even flavour-changing couplings to lepton pairs.  By proposing a specific mixing pattern in the lepton sector, we extracted the mixing parameter $\sin^2\theta = 0.37$ by accommodating $R_K = 0.745^{+0.090}_{-0.074}\pm 0.036$.  Using this information, we then made a prediction for the lepton non-universality in the $B\to K \tau \bar \tau$ and $K \mu\bar \mu$ decays as well as the lepton flavour violating  decays $B\to K (e \bar \tau, \tau \bar e)$.

\begin{acknowledgments}

C-W Chiang was supported in part by the MOST of ROC (Grant No.~MOST104-2628-M-008-004-MY4).  X-G He was supported in part by MOE Academic Excellent Program (Grant No.~102R891505) and MOST of ROC (Grant No.~MOST104-2112-M-002-015-MY3), and in part by NSFC (Grant Nos.~11175115 and 11575111) and Shanghai Science and Technology Commission (Grant No.~11DZ2260700) of PRC. G.~V. was supported in part by the DOE under Contract No. DE-SC0009974. G.~V. thanks the Physics Department of Jiao Tong University for their hospitality and partial support while this work was initiated. X.~G.~H. thanks Korea Institute for Advanced Study (KIAS) for their hospitality and partial support while this work was completed.

\end{acknowledgments}


\end{document}